\documentclass[aps,pre,preprint,superscriptaddress,showkeys]{revtex4}
\usepackage{graphicx}
\usepackage{times}
\begin{document}
\def\T{\Theta}
\def\D{\Delta}
\def\d{\delta}
\def\r{\rho}
\def\p{\pi}
\def\a{\alpha}
\def\g{\gamma}
\def\ra{\rightarrow}
\def\s{\sigma}
\def\b{\beta}
\def\e{\epsilon}
\def\G{\Gamma}
\def\om{\omega}
\def\pe{$1/r^\a$ }
\def\l{\lambda}
\def\f{\phi}
\def\w{\psi}
\def\m{\mu}
\def\t{\tau}
\def\c{\chi}

\title{Omnivory can both enhance and dampen perturbations  in food webs}
\author{Jaroslav Ispolatov \& Michael Doebeli \\
\vspace{-2mm}\normalsize Department of Zoology and Department of Mathematics, \\
\vspace{7mm}\normalsize University of British Columbia, Vancouver B.C., Canada
V6T 1Z4\\
{\it Corresponding author:} Michael Doebeli, Department of Zoology, 6270 University Boulevard, Vancouver, BC V6T 1Z4, Canada. email: doebeli@zoology.ubc.ca
}

\begin{abstract}
We investigate how perturbations propagate up and down a food chain with and without self-interaction and omnivory. A source of perturbation is a shift in death rate of a trophic level, and the measure of perturbation is the difference between the perturbed and unperturbed steady state populations. For Lotka-Volterra food chains with linear functional response, we show analytically that both intraspecific competition and intraguild predation can either dampen or enhance the propagation of perturbations, thus stabilizing or destabilizing the food web. The direction of the effect depend on the position of the source of perturbation, as well as on the position of the additional competitive and predatory links . These conclusions are confirmed numerically for a food chain with more realistic Type-II functional response. Our results support the positions of  both sides in the long-standing debate on the effect of intraspecific competition and omnivory on the stability of trophic systems.
\end{abstract}

\maketitle
\section{Introduction}
Food web stability is a topic with a tremendous scientific and practical significance \cite{pimm77,pimm02, quince05, dunne02, herendeen04, dunne06,
fowler02,  bascompte05} . A large number of both theoretical and experimental studies have addressed the question of how the stability of a food web depends on its structure, and specifically, on the presence of omnivory (see e.g \cite{pimm77, mccann97, diehl99,  holyoak98, rudolf07, holt97,fagan97,vandermeer06}, just to mention a few). 
Omnivory, defined broadly as feeding on more than one trophic level, which generally includes cannibalism, is a common phenomenon in natural communities \cite{rosenheim07, rudolf07}. However,  a consensus on the effect of omnivory on food web stability is still lacking. Thus,  according to some results from mathematical food web theory, omnivory destabilizes ecological communities \cite{pimm77, holt97}, whereas other models and experiments suggest that omnivory should be a strongly stabilizing factor in food webs \cite{mccann97, holyoak98, fagan97,hillerislambers06} .
The notion of  ``stability'' itself is broadly understood: the most common examples include a scarcity of secondary extinction events in response to removal or  a shift in birth or death rate of a species \cite{fowler02} , temporal characteristics of a return to a new equilibrium following such removal \cite{fagan97}, the mere existence of locally stable steady states
\cite{mccann97}, and a global permanency of food web dynamics \cite{law92,mccann97}. 

In addition to this apparent lack of consensus on how food web structure affects stability, it seems that there still exist fundamental questions related to the effect of omnivory on food web stability which have remained unanswered, and even more, not specifically formulated.  For example, in many practical situations a population at a certain position in a food web is harvested on a fairly permanent basis (for example, by fishing). Alternatively, reproductive rates or death rates of some species may shift due to environmental changes, endemics, etc.
Such long-term removal (or possibly, addition, \cite{diehl99}) of a resource from a certain trophic level naturally affects populations of the other levels in the food web. Generally, such removal may lead to extinction of certain species and to destabilization of initially steady state population dynamics in the food web. However, if the initial or unperturbed network is sufficiently stable, a continuous harvesting within certain limits may often result, after some transient behaviour, in constant shifts in population of other trophic levels, bringing the whole food web to a new steady state. The susceptibility of the steady state food web populations, or
the difference between the new and old, unperturbed steady states, is another characteristic of the stability of food webs with respect to shifts in death or birth rates or harvesting. This new characteristic has an evident practical significance. 
For example, it is interesting and important to determine ``sustainable''  levels of harvesting that produce a  limited  effect on the rest of the food web and on the harvested level itself. Intuitively it is clear that such a susceptibility-based definition of stability is related to previously used definitions, because a food web that is less susceptible to harvesting is less likely to leave a basin of attraction of a fixed point, a permanence domain, or suffer an extinction.

As with other definitions of stability,
susceptibility of a food web to  long-term  removal of resource from a certain trophic level depends on the position of this level in the web, predation strengths and natural death rates. The induced population shifts also depend on the structure (topology) of the food web, and specifically, on the presence, position, and strength of omnivory interactions.  This leads to the formulation of the main question we address in this paper:
How is the effect of the steady removal of a population of a certain level on the rest of the foodweb mediated by the presence of omnivory and self-interaction? In other words, we compare the steady state response of two model food webs, with and without omnivory trophic relations, to a shift of a death rate of a certain trophic level. 
When the response of the food web in the presence of omnivory is less pronounced than without it, omnivory is considered to be  a stabilizing factor. Conversely, when the induced changes in population densities s are greater with omnivory than without, omnivory is considered destabilizing.
To analyze this question in its most isolated and fundamental form, we focus on the topologically simplest form of a food web, a linear food chain. Also, our principle analysis deals with the  most basic form of predator-prey interactions, i.e., with Lotka-Volterra models with linear functional responses, which turn out  to be analytically tractable. Harvesting of a certain species is modeled as an introduction of a certain per capita rate of removal, which effectively modifies the death rate of the corresponding species in the food web..
We consider two possible forms of omnivory for linear food chains. First, we consider competition, or a negative self-interaction, at a certain trophic level, which  is modeled as an additional per capita death rate term proportional to the population density at this level. This can be thought of as omnivory because it is mathematically equivalent to cannibalism with a linear functional response. Second, we consider  intraguild predation, which we define  as  a ``shortcut'' trophic link introducing an additional predator-prey interaction between two non-adjacent trophic levels. 

Our conclusions confirm both existing viewpoints regarding stability and food web structure (and hence support neither as an overarching principle): both intraspecific competition and intraguild predation can either stabilize or destabilize a food chain. Whether stabilization or destabilization occurs depends on the relative position of the level at which the food web is perturbed on the one hand, and where the additional interaction links are introduced on the other hand. We also show numerically that both stabilizing and destabilizing effects of omnivory, derived analytically for Lotka-Volterra systems, qualitatively hold for food chains with more realistic, non-linear functional responses.  

The paper is organized as follows: In the next section we define the framework of our food web study, which is based on systems of coupled Lotka-Volterra equations, and find the steady state solution for the population densities at all levels of the food chain. 
Next we consider the effect on the steady state induced by a shift of a death rate of a given trophic level and determine how this perturbation spreads over the food chain.  We then derive how competition, or self-interaction, at a certain trophic level, modifies the steady state of a  food chain, and consider  the effect of such self-interaction on the spread of perturbation induced by a shift in a death rate. We show that depending on the position of self-interacting and death-rate modified levels,  self-interaction may inhibit or enhance such propagation. Then the same analysis is repeated for omnivory links, and  we present two examples of how such a link inhibits and enhances spread of perturbation. In the discussion, we present examples of quantitative similarities between the effects of omnivory and self-interactions in food chains with type-II functional responses food chain and our results for the Lotka-Volterra systems.   
We also illustrate how our results can be applied to more topologically complex food webs. The Appendix contains the description of a model with type-II functional responses in the presence of omnivory links. 

\section{Linear Lotka-Volterra chain}
We consider a linear food chain with nearest neighbour predator-prey interactions of Lotka-Volterra form. We do not limit ourselves to a specific number of levels in the chain and label the basal level by $0$ and the highest predator by $n$.  The rate of change of the population density $x_k(t)$ of trophic level $k$ (which will also be called ``species''  $k$) is given by
\begin{equation}
\label{lvii}
\frac{dx_k}{dt}=x_k\left[ {\l_ka_{k,k-1}x_{k-1}} - 
{a_{k+1,k} x_{k+1}}- d_k\right].
\end{equation}
Here $a_{i,j}$ determines the strength of predation  of species $i$ on species $j$, $\l_k$ is the conversion efficiency which connects the birth rate of species $k$ to the amount of other it consumes, and
$d_k$ is the per capita death rate of species $k$.
Evidently, the top predator, occupying the $n$th level, does not have any species that prey on it, which could be expressed by setting $x_{n+1}\equiv 0$.
In addition, the basal species, occupying the lowest (zeroes) trophic level, is characterized by
a logistic growth term, which mimics the finite input of energy into the system,
given in the form of a finite carrying capacity $K$ for the basal species.
Without loss of generality, the linear death term for the basal species can be absorbed into the linear part of the birth term $\b$, which yields the following for the rate of change of the population density of the basal species:
\begin{equation}
\label{lv0}
\frac{dx_0}{dt}=x_0\left[\b\left(1-\frac{x_0}{K}\right) -  {a_{1,0}x_{1}}\right].
\end{equation}

\begin{figure}
\includegraphics[width=.4\textwidth]{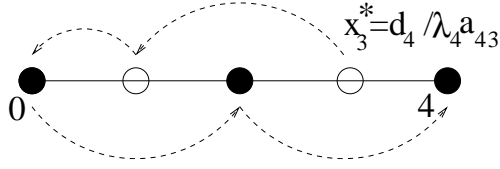}
\caption{\label{lv}
Direction of recurrence for finding the steady state populations $x_k^*$ in a 5-level
(0 -- 4) Lotka-Volterra linear chain. The solution starts with finding the population of the second from the top level, $x_3^*$ and is propagated downwards so that $x_1^*$ is determined via (\ref{steady_g}). Then $x_0^*$ is determined using the Eq.~(\ref{boundary_bottom}), and the recursion is propagated upward to determine $x_2^*$ and $x_{4}^*$. Here and below the levels of $\{n\}$  ( or ascending) partition are shown as black circles and the levels belonging to the $\{n-1\}$  (or descending) partition are shown as white circles.}
\end{figure}

We look for a steady state solution of this system, $d\vec x^* / dt=0$,
which is defined by the following system of linear equations:
\begin{equation}
\label {steady_g}
 x_{k-1}^* = x_{k+1}^*  \frac{a_{k+1,k}}{\l_k a_{k,k-1}} + \frac{d_k}{\l_k a_{k,k-1}}
\end {equation}
Recurrence relations  of the form $x_k^*= F(x_{k+2}^*)$ given by Eqs.~(\ref{steady_g})  indicate that in the steady state the Lotka-Volterra food chain splits into two subsets or partitions: the steady state of even-numbered levels depend only on another even-number level and a set of constants, and  similarly, odd-number levels are coupled only to odd-number levels. To solve the system  (\ref{steady_g}) one needs to consider the closing equations which describe the steady state populations of the top and bottom levels. Because the dynamics of the top predator population depends only on  $x_{n-1}$,
the first closing equation (or the boundary condition for the recurrence relation (\ref{steady_g})) 
defines the population of the $n-1$ level,
\begin{equation}
\label{boundary_top}
x_{n-1}^*=\frac{d_n}{\l_n a_{n,n-1}}.
\end{equation}
Consequently, sequential application of  recurrence (\ref{steady_g}) allows one to determine the  populations of lower levels of the same parity as $n-1$, down to level one or zero. 
The second boundary condition, 
\begin{equation}
\label{boundary_bottom}
 x_1^*=\frac{\b}{a_{1,0}}\left( 1- \frac{x_0^*}{K}\right),
\end{equation}
which follows from the stationarity of the basal population, connects the populations of levels one and zero, thus linking the even and odd partitions. Finally, applying  (\ref{steady_g}) upstream, or
expressing $x_{k+1}^*$ through $x_{k-1}^*$, we determine the steady state populations of levels of the same parity as $n$. Thus Eqs.~(\ref{steady_g}, \ref{boundary_top}, \ref{boundary_bottom}) define the stationary solution of the whole Lotka-Volterra food chain:
The recurrence starts from the $n-1$th level, descends to level one or zero depending on the parity of $n$, and then ascends to the top predator level $n$. This is illustrated schematically  in Fig.~\ref{lv}.

In the following we will widely use this natural separation of the  food chain into two partitions:
The regression in the ``descending'', or ``$\{n-1\}$ '' partition   starts from the $n-1$st level  and goes down along the indexes of the same parity as $n-1$ to level one or zero, depending on whether $n$ is even or odd. It is important to note that the steady state population of a level of the descending partition is determined only by the level immediately above it in the same partition. 
Conversely, the regression in the``ascending'',  or ``$\{n\}$ '' partition, which spans indexes of the same parity as $n$, starts from zero or one and goes up to $n$.  The population density of a given level in the ascending partition depends only on the population of its lower nearest neighbour in the same partition.

Assuming for simplicity equality of the predation intensities, $a_{ij}=a$, conversion coefficients, $\l_i=\l$, and death rates, $d_i=d$, it is possible to express the general solution for the steady state populations in a compact form.
The system of equations for the steady state populations (\ref{steady_g}) becomes
\begin{equation}
\label {steady}
\l x_{k-1}^* =x_{k+1}^* + \frac{d}{a}.
\end {equation}
The notation
\begin{equation}
\label{y}
y_k=x_k^* \l^{-k/2} a/d,
\end{equation}
reduces the recurrence relation (\ref{steady}) to
$$
y_{k-1}=y_{k+1} + \l^{-(k+1)/2}.
$$
Thus, $y_{k}$ can be expressed as a sum of a geometric progression,
and the general solution for arbitrary $i$ and $k$ reads 
\begin{equation}
\label{sol}
x_{k+2i}^*=\l^{i}  x_k^* - \frac{d}{a} \;\frac{1 -\l^{i}}{1-\l}.
\end{equation}
The first boundary condition defines the population of the $n-1$ level,
\begin{equation}
\label{1b}
x_{n-1}^*=\frac{d}{\l a},
\end{equation}
and, via (\ref{sol}), the populations of the levels of the $\{n-1\}$  partition.
The second boundary condition links $\{n\}$   to $\{n-1\}$  partitions,
\begin{equation}
\label{2b}
 x_1^*=\frac{\b}{a}\left( 1- \frac{x_0^*}{K}\right),
\end{equation}
and yields the full explicit form of the stationary state populations. 
For example, for odd $n$,
\begin{eqnarray}
\label{full_solution}
\nonumber
x_{2k}^*=\frac{d}{a} \;\frac{\l^{k-(n+1)/2} -1}{1-\l}\\
x_{2k+1}^*=\frac{\b \l ^k} {a} \left(1-\frac{x_0}{K}\right) -\frac{d}{a}\;\frac {1- \l^k}{1-\l}
\end{eqnarray}
In the following we denote, where possible,  the levels of the $\{n-1\}$  partition as $i$ and levels of the $\{n\}$  partition as $j$.
It also follows from (\ref{full_solution}) that for a given set of rate constants, the maximum length of the chain is finite and limited by the requirements that the descending partition does not exceed the carrying capacity even at its maximum, $x_0^*<K$ and the ascending partition does remain positive even at its minimum, $x_n^*>0$.

\section{Perturbations caused by shifts in the death rate of a single level}
Now we consider how changes of conditions  at a certain trophic level  affect the steady state populations of all levels of the Lotka-Volterra food chain. We assume  the simplest form of a perturbation equivalent to imposing a certain probability per unit time $\e a$ for each individual inhabiting the level $p$ to be removed from  the population. This can be taken into account by modifying the death rate coefficient, $d_p'=d_p+\e a$
(the factor $a$ is introduced for further convenience).

We now investigate how such a shift in death rate of the level $p$ affects the steady state population of all levels, or, in other words, how the perturbed steady state populations $x_k'$ are different from the unperturbed ones $x_k^*$.  For simplicity in the following we will assume that predation intensities  and conversion coefficients are constant for all levels, $a_{ij}=a$ and $\l_i=\l$.
Evidently, the shift in a death rate affects only  the levels in the direction of recurrence from the perturbed site $p$. That is, if a level of  the $\{n\}$ partition is perturbed (so that  $p \in \{n\}$), the perturbation first manifests itself at the $p-1$ level,   while $x_{p+1}^*$ and populations of higher levels of the $\{n-1\}$  partition remain unaffected. It follows from Eq.~(\ref{steady}) that
\begin{equation}
\label{perturbation}
\l x_{p-1}'=x_{p+1}^*  +\frac{d_p} { a} + \e,
\end{equation}
and
\begin{equation}
\label{perturbation2}
x_{p-1}'=x_{p-1}^* + \frac {\e} {\l}.
\end{equation}
Consequently, the perturbation propagates down the food chain to the zeroth level if $n$ is odd and to first level if $n$ is even,
\begin{equation}
\label{p1}
 x_{i}'=x_{i}^*, \; i>p; \;\;  x_{i}'=x_{i}^* + \frac{\e}{\l^{(p+1-i)/2}},  \; i<p. 
\end{equation}
Then at the basal level, due to the second boundary condition (\ref{boundary_bottom}), the perturbation changes sign,
 \begin{equation}
\label{p2}
 x_1'=x_1^* - \frac{b}{aK} \frac{\e}{\l^{(p+1)/2}}
\end{equation}
for odd $p$ and $n$, and
\begin{equation}
\label{p3}
 x_0'=x_0^* - \frac{aK} {b} \frac{\e}{\l^{p/2}}
\end{equation}
for even $p$ and $n$.
For the $\{n\}$  partition the ``ascending'' regression yields
\begin{equation}
\label{p5}
 x_{j}'=x_j^* - \g^{\psi} \frac{\e}{\l^{(p-j+1)/2}}
\end{equation}
Here we introduced a universal notation for odd and even $n$,
\begin{equation}
\label{g}
\g \equiv \frac{b}{a\sqrt{\l}K};
\end{equation}
with $\psi=1$ for odd $n$ and  $\psi=-1$ for even $n$. Thus, when the death rate of a species in $\{n\}$ partition is increased, the population of the levels of $\{n-1\}$ partition lying below $p$ increases, while the population of all levels of the  $\{n\}$ partition, including the perturbed level $p$, decreases.

Similarly, when the death rate of the level $p \in \{n-1\}$ is shifted, the perturbation propagates recurrently upwards  from $j=p+1$,
\begin{equation}
\label{p6}
x_i'=x_i^*,\: x_j'=x_j^*, \; j<p, \;\; x_{j}'=x_j^* -{\e \l^k}, \; j=p+1+2k.
\end{equation}
In this case the $\{n-1\}$ partition, including the directly affected level $p$ remains unperturbed, while the population of the levels of $\{n\}$n partition that are above $p$ is decreased.

\begin{figure}
\includegraphics[width=.7\textwidth]{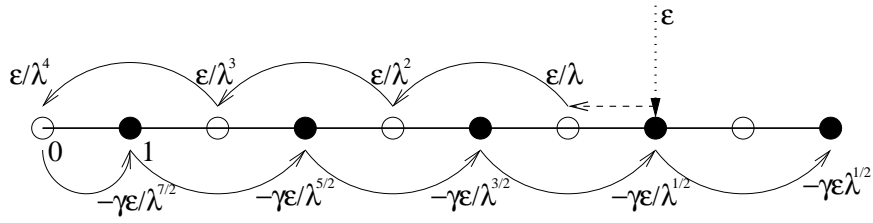}
\caption{\label{pertfig}
Propagation of perturbation caused by a shift by $\e a$ of the death rate of a level belonging to the $\{n\}$ partition; the recurrence relation (\ref{perturbation}), perturbed by the shift in death rate, is shown by a dashed arc. The resulting spread of perturbation is shown by solid line arcs. The numbers above and below vertices indicate the change in the population of the corresponding trophic level induced by the perturbation.}
\end{figure}

\begin{figure}
\includegraphics[width=.7\textwidth]{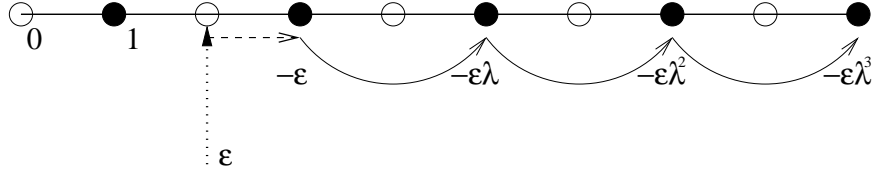}
\caption{\label{pertfig_up}
Propagation of perturbation caused by a shift by $\e a$ of the death rate of  a level belonging to the $\{n-1\}$ partition;  notations and symbols are the same as in Fig.~\ref{pertfig}. }
\end{figure}

We illustrate the propagation of perturbations in a linear Lotka-Volterra food chain in Figs.~\ref{pertfig}, \ref{pertfig_up}.

\section{Self-interaction}

Self-interaction at level $s$ in a linear food chain is taken into account by adding $-\a a x_s^2$ to the right-hand side of the  rate equation for $x_s$,
\begin{equation}
\label {si1}
\frac{dx_s}{dt}=x_s\left[ {\l a x_{s-1}} - 
{a  x_{s+1}}- d   - \a a x_s\right].
\end{equation}
Here $\a$ defines the intensity of self-interaction, relating it to the intensity of predation $a$. As mentioned, the self-interaction can be thought of as either competition or cannibalism. As a consequence of self-interaction, the $s$th  equation for the steady state population is modified to
\begin{equation}
\label {steady_si}
\l x_{s-1} =x_{s+1} + \frac{d}{a} +\a x_s.
\end {equation}
Thus, similarly to the case of death rate shift,  the self-interaction of a level $s$ of one partition affects the population of those levels of the opposite partition which are recursively downstream from the perturbation site, that is, $s-1$ if $s \in \{n\}$ and 
$s+1$ if $s \in \{n-1\}$.
We denote $\d x_k$ the change to the steady state population induced by a self-interaction, so that the new populations in system with self-interaction are $x_k^*+ \d x_k$. Then, from (\ref{si1}),
\begin{equation}
\label {corr_si}
\l \d x_{s-1} = \d x_{s+1}  +\a (x_s^* + \d x_s).
\end {equation}
The unperturbed steady state populations $x_k^*$ are given by (\ref{full_solution}).

There are two principal cases:
\begin{itemize}
\item $s \in \{n\}$. Here self-interaction directly affects the $s-1$ level, 
\begin{equation}
\label {rs_si}
 \l \d x_{s-1} = \a (x_s^* + \d x_s), 
\end{equation}
and recursively affects all lower levels $i<s-1$ of the $\{n-1\}$ partition,
\begin{equation}
\label {ri_si}
 \l \d x_{i-2} = \d x_i
\end{equation}
 and  all levels $j$ of the $\{n\}$ partition,
\begin{equation}
\label {rj_si}
 \l \d x_{j} = \d x_{j+2} 
\end{equation}
From the boundary condition (\ref{2b}) and regressions (\ref{ri_si}, \ref{rj_si}) it follows that
\begin{equation}
\label{sw}
 \d x_{s}=-\g^{\psi}\sqrt{\l} \d x_{s-1}.
\end{equation}
Thus 
\begin{equation}
\label{sii}
\d x_i = \frac { \a x_s^*} {( 1+ \a  \g^{\psi}/\sqrt{\l}) \l^{(s+1-i)/2}}
\end{equation}
for $i<s$, and 
\begin{equation}
\label{sij}
\d x_j = - \g^{\psi} \frac { \a x_s^*} { (1+ \a^*) \l^{(s+1-j)/2}}
\end{equation}
for any $j$. We denote by $\a^* \equiv \a \g^{\psi}/\sqrt{\l} $ a frequently occurring group of parameters which is proportional to $\a$.

\item  $s \in \{n-1\}$. Here self-interaction directly affects the $s+1$ level and recursively affects only the higher levels of the same $\{n\}$ partition. Hence
\begin{equation}
\label{sij2}
\d x_j = - \a \l^{(j-s-1)/2}x_s^*, \;  j>s
\end{equation}
\end{itemize}

Thus a self-interaction in $\{n\}$ partition increases the population of the lower levels of the $\{n-1\}$ partition but decreases the population of all levels in $\{n\}$ partition.  A self-interaction in $\{n-1\}$ partition decreases the population of the higher levels of $\{n\}$ partition and does not affect the $\{n-1\}$ partition.

\section{Death rate shifts with self-interactions}
Having established how perturbations induced by a shift in a death rate propagate in the  linear Lotka-Volterra chain and how the linear chain is affected by self-interaction,  we now consider the first of our two principal questions: How does self-interaction affect the propagation of perturbations?
Specifically, we would like to know if self-interaction stabilizes the system by reducing the amplitude of perturbations induced  by a shift of death rate, or, on the contrary, whether self-interaction destabilizes the system by increasing this amplitude. 
 We denote by  $\D x_k$ the change in the steady state population of level $k$ resulting from the death rate shift
$d_p'=d + \e a$  in  a chain with self-interaction of the form given by the Eq.~(\ref{si1}) at the level $s$ . We are interested only in the terms that vanish when the death rate shift is zero, thus $\D x_k$ does not contain $\e$-independent terms proportional to $x_s^*$.
Formally, the recurrent relation for $\D x_k$ follows from Eqs.~(\ref{perturbation}, \ref{corr_si}),
\begin{equation}
\label{Dx}
\l \D x_{k-1} = \D x_{k+1}+ \d_{k,p} \e + \d_{k,s} \a \D x_k,
\end{equation}
where Kronecker's delta-symbol $\d_{i,j}=1$ for $i=j$ and  $\d_{i,j}=0$ when $i\ne j$.
Since  death rate shifts and self-interaction can occur at levels belonging to any of the two partitions, there are 4 classes of scenarios. It turns out that among these 4 scenarios there are two examples of a stabilizing effect of a self-interaction, one of a destabilizing effect, and one of an absence of any effect on stability. 
\begin{itemize}
\item Both the perturbed level $p$ and the self-interaction level $s$ are located in the ascending $\{n\}$ partition, Fig.~\ref{si1fig}.
\begin{figure}
\includegraphics[width=.7\textwidth]{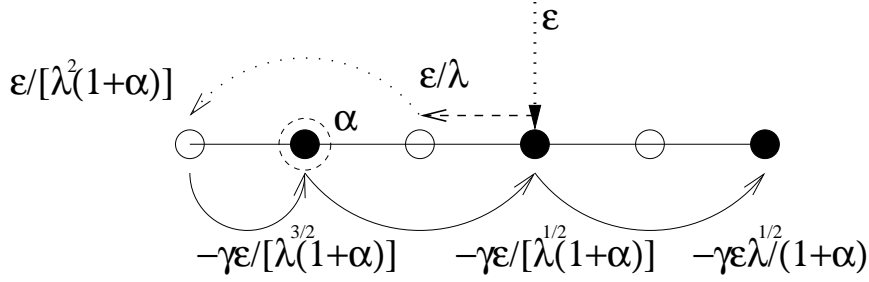}
\caption{\label{si1fig}
The effect of self-interaction at level 1 (shown by a large dashed circle) on propagation of perturbation caused by a shift by $\e a$ of the death rate at level 3; both self-interacting and perturbed levels belong to the $\{n\}$ partition. The recurrence relation modified by self-interaction (\ref{steady_si}) is shown by a dotted arc}
\end{figure}
First consider the case when $p>s$.  For $i>s$ the levels experience only the effect of the death rate perturbation and 
$\D x_i$ is given by (\ref{p1}). For the $s-1$st  level Eq.~(\ref{Dx}) reads 
\begin{equation}
\label{Dx1}
\l \D x_{s-1} = \D x_{s+1} +  \a \D x_s,
\end{equation}
where $\D x_{s+1} ={\e}/{\l^{(p+s)/2}}$ from  (\ref{p1}).
Similarly to  (\ref{sii}, \ref{sij}), we express $\D x_s$ through $\D x_{s-1}$ and obtain for $i<s$
\begin{equation}
\label{Dx2}
\D x_{i}=
\frac {\e} {( 1+ \a^*) \l^{(p+1-i)/2}},
\end{equation}
and 
\begin{equation}
\label{Dx3}
\D x_j = - \g^{\psi} \frac {\e} { (1+ \a^*) \l^{(p+1-j)/2}}
\end{equation}
for any $j$.
Now we consider the case when $p<s$. Then self-interaction enters the recurrence relation first,
\begin{equation}
\label{x1}
\l \D x_{s-1}= \a \D x_s.
\end{equation}
The equation for the level with a shifted death rate becomes
\begin{equation}
\label{x11}
\l \D x_{p-1}= \frac{ \a \D x_s}{\l^{(s-p)/2} +\e}
\end{equation}
Expressing $\D x_s$ through $\D x_{p-1}$ using the boundary condition at the base of the chain,
we obtain
 \begin{equation}
\label{x111}
 \D x_{p-1}= \frac{ \e}{(1+\a^*) \l }
\end{equation} 
Consequently, the results for $i<p$ and all $j$ are given by
Eqs.~(\ref{Dx2},\ref{Dx3}).
Thus we observe that when both the perturbed and self-interaction levels belong to the $\{n\}$ partition, self-interaction weakens the effect of death rate shift on other levels, and therefore has a stabilizing effect.
\item The perturbed level $p$ belongs to the ascending $\{n\}$ partition and the level of self-interaction $s$ belongs to the descending $\{n-1\}$ partition, Fig.~\ref{si2fig}. 
\begin{figure}
\includegraphics[width=.7\textwidth]{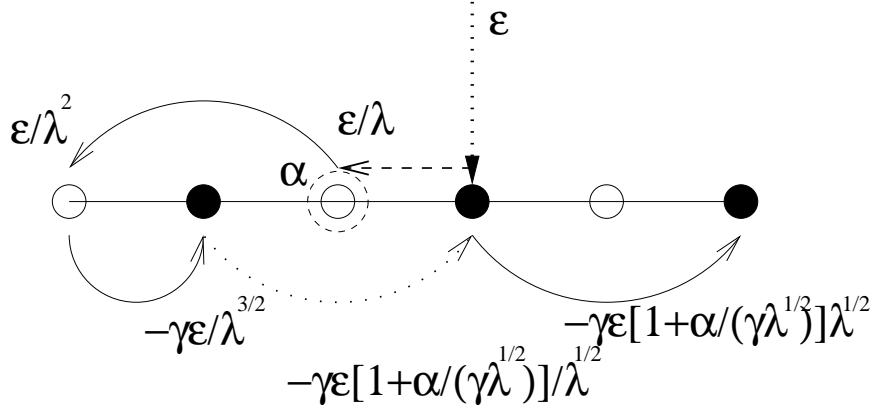}
\caption{\label{si2fig}
The effect of self-interaction at level 2  of  the $\{n-1\}$ partition with a perturbation caused by a shift  of the death rate at level 3  of the $\{n\}$ partition.}
\end{figure}
If $p<s$, no joint effect occurs as the self-interacting level remains unperturbed by the death rate shift.
However, in the opposite case of $p>s$,  $\D x_i= {\e}/{\l^{(p+1-i)/2}},  \; i<p.$,  see (\ref{p1}),
and $\D x_j= - \g^{\psi} {\e}/{\l^{(p-j+1)/2}},\; j<s$, see (\ref{p5}).
For $j>s$ the effect of self-interaction simply adds to the effect of perturbation,
\begin{equation}
\label{ps2}
\D x_j =  - \g^{\psi}\frac{\e} {\l^{(p-j+1)/2}} \left (1+\frac {\a} {\g^{\psi} \l^{1/2}} \right).
\end{equation}
Here we observe that self-interaction enhances the effect of perturbation, and hence is destabilizing. 
\item The perturbed level  $p$ belongs to the $\{n-1\}$ partition and the level of self-interaction $s$ belongs to the $\{n\}$ partition, Fig.~\ref{si3fig}.
There is no joint effect when $p>s$. When $p<s$, the correction to the $j>p$ levels of $\{n\}$ partition, including the level $s$, comes both self-interaction and shift of death rate.
\begin{figure}
\includegraphics[width=.7\textwidth]{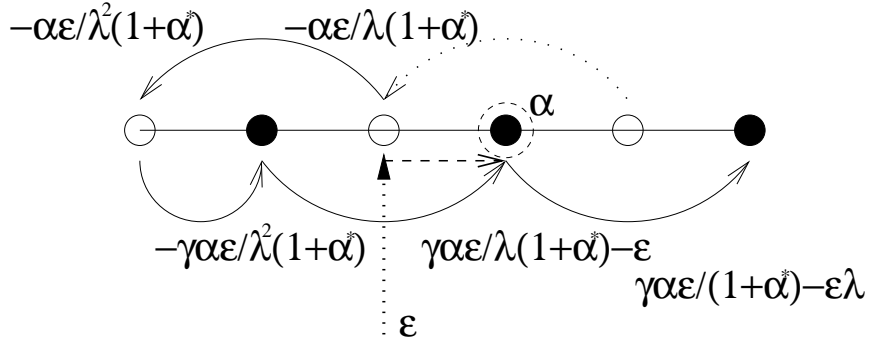}
\caption{\label{si3fig}
The effect of self-interaction at level 3 of the $\{n\}$ partition when perturbation is caused by a shift of the death rate at level 2 of the $\{n-1\}$ partition.}
\end{figure}
Hence, similarly to (\ref{sii}), we write for the $i<s$ levels of $\{n-1\}$ partition  
\begin{equation}
\label{p31}
\D x_i = - \frac {\a  \e } {1+ \a^*} \l^{(i-p)/2-1},
\end{equation}
and, similarly to (\ref{sij}),  for $j<p$ levels of $\{n\}$ partition   
\begin{equation}
\label{p32}
\D x_j =  \g^{\psi} \frac {\a  \e } { 1+ \a^*} \l^{(j-p)/2-1}.
\end{equation}
For $j>p$, a direct correction from the death rate shift is added,
\begin{equation}
\label{p33}
\D x_j =  \g^{\psi} \frac {\a  \e } { 1+ \a^*} \l^{(j-p)/2-1} - \e \l^{(j-p-1)/2}.
\end{equation}
Thus the self-interaction dampens the propagation of perturbations and stabilizes the system.
\item Finally,  both the perturbed level $p$ and self-interaction level $s$ are located in the descending ($\{n-1\}$ ) partition, Fig.~\ref{si4fig}. In this case no joint effect of  perturbation and self-interaction occurs since the descending partition levels remain unperturbed and do not add any terms depending on $\e$ through self-interaction.
\begin{figure}
\includegraphics[width=.7\textwidth]{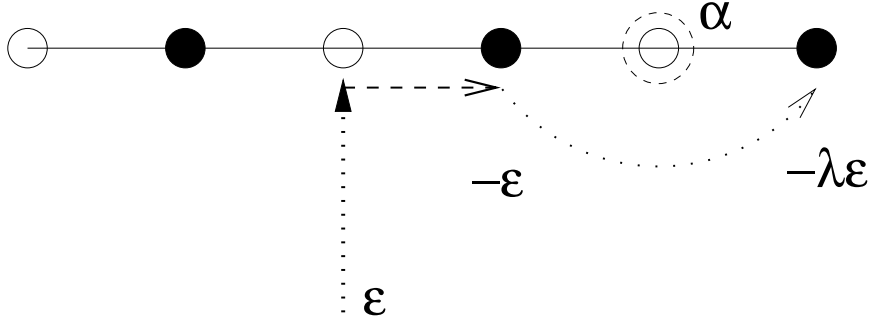}
\caption{\label{si4fig}
The effect of self-interaction at level 3 of the $\{n\}$ partition when perturbation is caused by a shift of the death rate at level 2 of the $\{n-1\}$ partition..}
\end{figure}
\end{itemize}

A qualitative summary of the effect of self-interaction on propagation of perturbations is presented in the Table~1.
\begin{table}
\centering  \begin{tabular} { c | c | c }
    \hline
     & $s \in \{n\}$ & $s \in \{n-1\}$\\ \hline
 $p \in \{n\}$    &  dampens  & enhances \\  \hline
 $p \in \{n-1\}$    &  dampens & no effect
\\  \hline
  \end{tabular}
\end{table}

\section{Omnivory shortcuts} \label{os}
Here  we address the second principal question of our study: How does an intraguild trophic relation, or  a predator-prey interaction connecting two non-adjacent trophic levels, affect
the spread of perturbations in the food chain (where we again consider perturbation induced by shifting the $p$-level death rate)? A shortcut  trophic link connects what we call an upper level $u$ to a lower level $w$, $u-w \ge 2$. As in the self-interaction case, we would like to know if such an omnivory link stabilizes the system by reducing the perturbations induced by shifts in death rates.
The shortcut  link modifies the rate equations for the $u$ level by adding a gain term
$$
\a \c a x_w x_u
$$
and for the $w$ level by adding a loss term
$$ 
 -\a  a x_w x_ u.
$$ 
As in the case of self-interaction, the dimensionless factor $\a$ defines the intensity
of  a shortcut relative to regular predation. 
The gain term has a conversion factor $\c$ which is distinct from $\l$ since preying across several trophic level can have different efficiency in terms of the birth rate of the predator, compared to preying on the next nearest level below. 
The $u$th and $w$th equations for the stationary concentrations are  modified accordingly,
\begin{eqnarray}
\label {shLV}
\nonumber
\l x_{u-1} =x_{u+1} + \frac{d}{a} -\c \a x{_w} ,\\
\l x_{w-1} =x_{w+1} + \frac{d}{a} + \a x_u
\end{eqnarray}
As in the case of self-interactions, it follows that the omnivory link between $u$ and $w$ directly affects the population of the levels that are recursively downstream from $u $ and $w$. Similarly to the self-interaction case, we denote corrections to the population steady states resulting from the  effect of a death rate shift on a food chain with shortcuts by $\D x_k$. A formal recursion equation for  $\D x_k$, analogous to Eq.~(\ref{Dx}) reads
\begin{equation}
\label{Dxs}
\l \D x_{k-1} = \D x_{k+1}+ \d_{k,p} \e + \d_{k,w} \a \D x_u - \d_{k,u} \c \a \D x_w 
\end{equation}
Since the perturbed level, the upper level of a shortcut, and the lower level of the shortcut can belong to either $\{n\}$   or $\{n-1\}$ partition, there exist $2^3 =8$ distinct scenarios.
Below we present detailed analysis for two examples of  enhancement and dampening of perturbations by a shortcut. The analysis of the remaining  6 scenarios is straightforward and yields additional stabilizing and destabilizing examples.

 \subsection{Example of destabilization} \label{ed}
\begin{figure}
\includegraphics[width=.7\textwidth]{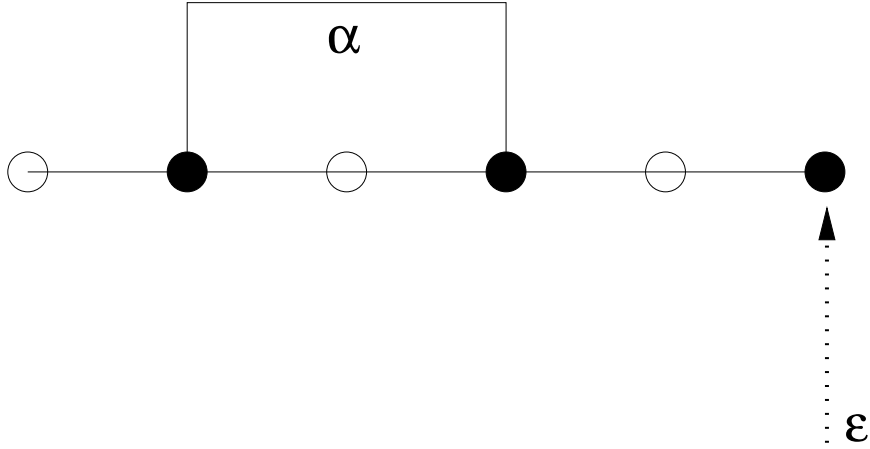}
\caption{\label{sh1fig}
The effect of omnivory interaction between levels $u=3$ and $w=1$  on the propagation of a perturbation caused by a shift by $\e a$ of the death rate at level $p=5$; all these 3 levels $\{u,w,p\}$  are in $\{n\}$ partition.}
\end{figure}

 We consider a case where all 3 levels $p,u,w$ belong to the $\{n\}$ partition, as illustrated in Fig.~\ref{sh1fig}.
Let us first assume that the death rate shift occurred at $p>u$. 
The corrections to population induced by the omnivory link depend on $\D x_u$ and $\D x_w$, which, in their turn, could be expressed through $x_{w-1}$ via the relation (\ref{sw}) linking both partitions.
\begin{equation}
\label{es1}
\D x_w = - \g^{\psi}  \sqrt{\l} \D x_{w-1}; \; \D x_u = \l^{(u-w)/2} \D x_w  . 
\end{equation}
On the other hand, $\D x_{w-1}$ depends on $\D x_w$ directly (see Eq.~(\ref{Dxs}) for $k=w$), and 
on $\D x_{u-1}$ recurrently through $\D x_{w+1}$. $\D x_{u-1}$ in turn  directly depends on $\D x_u$ via Eq.~(\ref{Dxs}) for $k=u$, and recurrently through $D x_{u+1}$, which depends on the death rate shift at $p$.
Putting together all these dependencies yields an equation for $\D x_{w-1}$ and $\D x_{w}$,
\begin{equation}
\label{es2}
\D x_{w-1} = \e \l ^{(w-p-2)/2} + \frac {\D x_w \a} {\l} \left( \l^{(u-w)/2} - \c \l^{(w-u)/2} \right). 
\end{equation}
Combining Eqs.~(\ref{es1}, \ref{es2}) produces an expression for the  population correction  $i<w$,
 \begin{equation}
\label{es3}
\D x_{i} = \frac{\e \l ^{(i-p-1)/2}}
{1 + \a^*  \left( \l^{(u-w)/2} - \c \l^{(w-u)/2} \right)},
\end{equation}
and, via (\ref{sw}) , for any $j$,
\begin{equation}
\label{es4}
\D x_{j} = -\g^{\psi} \frac{\e \l ^{(j-p-1)/2}}
{1 + \a^*  \left( \l^{(u-w)/2} - \c \l^{(w-u)/2} \right)},
\end{equation}
For $w<i<u$, only the $u$ and $p$ levels affect the population perturbation,
so that,
\begin{equation}
\label{es5}
\D x_{i} = \frac{\e \l ^{(i-p-1)/2}}
{1 - \a^* \c \l^{(w-u)/2}},
\end{equation}
and for $i>u$ no effect of the omnivory link is present. 
These results hold for arbitrary positions of the death-rate-shifted level $p$ with respect to $u$ and $w$:  Once $p \in \{n\}$, all $\{n\}$ partition levels, including $u$ and $w$, are perturbed. The only difference with the above case occurs when $p<i$. In this  case, the direct effect of death-rate perturbation $\e \l ^{(i-p-1)/2}$ should be subtracted from $\D x_i$ defined via Eqs.~(\ref{es3}, \ref{es5}).
It follows from Eqs.~(\ref{es3}, \ref{es4}) that the sign of the effect depends on the values of $\l$ and $\c$.
In a common scenario with  $\l <1$ and $\c \approx \l$, the effect of the omnivory link at the $i<w$ and all $j$ levels is destabilizing: the perturbation induced by a  shift in death rate is enhanced  in the presence of the omnivory  link. However the strongest enhancement or destabilization, which is universal and manifests itself for arbitrary $\l$ and $\c$,  occurs for the $w<i<u$ levels of the  $\{n-1\}$ partition, Eq.~(\ref{es5}). In the idealized case with $\l=\c=1$, the two terms in brackets in the denominators of (\ref{es3}, \ref{es4}) cancel each other, so that the destabilizing effect of the omnivory can be seen only at the levels with $w<i<u$.

\subsection{ Example of stabilization} \label{es}
\begin{figure}
\includegraphics[width=.7\textwidth]{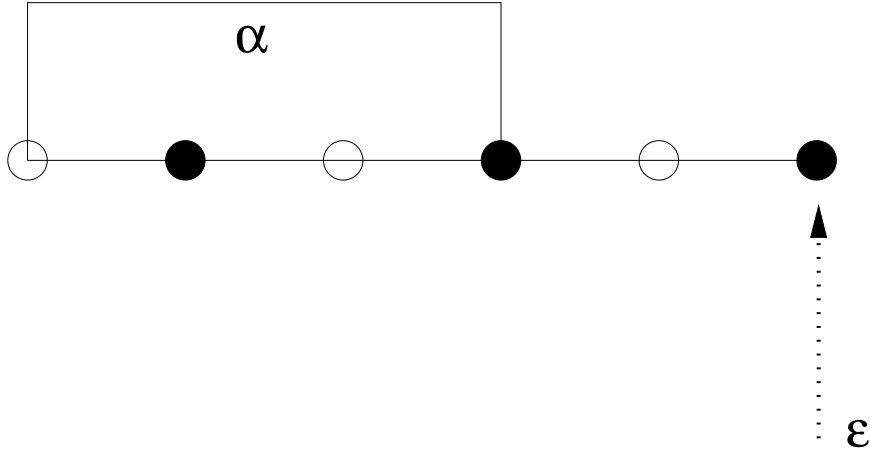}
\caption{\label{sh2fig}
The effect of the omnivory interaction between levels $u=3$ of the $\{n\}$ partition and $w=0$ of the  $\{n-1\}$ partition on propagation of perturbation caused by a shift by $\e a$ of the death rate at level $p=5$, which belongs to the $\{n\}$ partition.}
\end{figure}

 Here we consider a case where the omnivory link connects a level $u$ of the $\{n\}$ partition to a level $w$ of the $\{n-1\}$ partition,
 while the death rate perturbation occurs at a level $p$ of the $\{n\}$  partition, as illustrated in Fig.~\ref{sh2fig}.
First consider the case when $p>u$. From (\ref{Dxs}) it follows that 
\begin{equation}
\label{es6}
\D x_{u-1} = \e \l ^{uj-p-2)/2} - \D x_{w} \a \c /\l
\end{equation}
Taking into account that $\D x_w$ is related to $ \D x_{u-1}$ via a simple recurrence, 
$$
\D x_w = \D x_{u-1} \l^{(w-u+1)/2},
$$
one obtains for $i<u$
\begin{equation}
\label{es7}
\D x_i = \frac { \e \l^{i-p-1)/2}}{1+ \a \c \l^{(w-u-1)/2}}
\end{equation}
Consequently, for $j<w$ the perturbation is a simple recurrent continuation of (\ref{es7}),
\begin{equation}
\label{es8}
\D x_j = - \frac { \e \g^{\psi} \l^{i-p-1)/2}}{1+ \a \c \l^{(w-u-1)/2}}.
\end{equation}
For $\D x_{w+1}$ it follows from   (\ref{shLV}) that
\begin{equation}
\label{es99}
\l \D x_{w-1} = \D x_{w+1} + \a \D x_u = \D x_{w+1} ( 1+ \a \l^{(u-w-1)/2}).
\end{equation}
Hence for $j>w$
\begin{equation}
\label{es9}
\D x_{j} = - \frac{ \e \g^{\psi} \l^{j-p-1)/2}}{\left(1 + \a \c \l^{(w-u-1)/2}\right) \left( 1+ \a \l^{(u-w-1)/2}\right)}.
\end{equation}
A similar scenario occurs when $w<p<u$; to get the population corrections one needs to subtract the direct effect of the death rate shift $\e \l^{(i-p-1)/2}$ from the corrections $\D x_{i}$ given by
Eq.~(\ref{es7}) for $i>p$.
When $p<w$, there is no effect of the shortcut on the propagation of perturbation in the $\{n-1\}$ partition and the lower part, $j<w$, of the $\{n\}$ partition. However, for $j>w$ the shortcut does have an effect on perturbation amplitude, which is expressed by the second term in the denominator of (\ref{es9}),
\begin{equation}
\label{es10}
\D x_{j} = - \frac{ \e \g^{\psi} \l^{j-p-1)/2}}{ 1+ \a \l^{(u-w-1)/2}}
\end{equation}

Thus, we observe that an omnivory link between  $u \in \{n\}$ and $w \in \{n-1\}$ stabilizes the food chain by reducing the effect of a death rate shift on the population. This conclusion holds for arbitrary $\l$ and $\c$, and for any relative position of the levels linked by omnivory and the level at which the death rate perturbation occurs. 

\section{Discussion and Conclusions}

Using linear Lotka-Volterra food chains, we have investigated how perturbation at one trophic level cascades to other trophic levels, and how such perturbation cascades are affected by the introduction of additional links in the food chain, representing either self-interaction (intra-specific competition or cannibalism), or omnivory. Our results show that such additional links can either dampen or enhance the perturbation cascade, depending on the position of the level at which the perturbation occurs, and the levels connected by the additional link. Thus, self-interaction and omnivory can be either stabilizing or destabilizing, depending on the details of the setup. 

In establishing these results, we obtained a closed-form solution for the steady state populations of Lotka-Volterra food chains of arbitrary length, and we
derived analytic expressions for the changes in the steady states in response to 
shifts of death rate, self-interaction and additional intraguild trophic relations.  

Our definition of stability, based on susceptibility of steady state populations to 
long-term removal or addition of a population of a certain level
 is somewhat different from previously considered ones.
These previously considered definitions of stability include the size of cascading extinction events following the removal of randomly chosen species \cite{fowler02},  permanency, or localization to a certain  region of phase space for systems with possibly non-stationary asymptotic behaviour \cite{law92},  time of relaxation to a new equilibrium \cite{fagan97},
and the mere convergence to a steady state \cite{mccann97}.  Our definition is inspired by practical applications in which the effect of continuous harvesting of a certain species on the whole food web needs to be considered. 
However, susceptibility to perturbation is intrinsically related to other definitions of stability, such as the existence of steady states and lack of extinction events. In response to a perturbation, a less susceptible (and thus more stable, according to our definition) system deviates 
less from  a steady state than a more susceptible one, thus having a lower probability of leaving the basin of attraction of a stable fixed point, or cross the boundary of a phase region with permanent behaviour.  This can be illustrated by considering, for example the expressions for stabilizing (\ref{Dx3}) and destabilizing (\ref{ps2}) effects of self-interaction.
When $x_j + \D x_j$ is negative, a steady state is no longer possible, which means that either the population of the level $j$(and consequently, all higher levels)  becomes extinct, or the behaviour of the system becomes non-stationary. When the effect of additional links is stabilizing,
 (\ref{Dx3}), this happens for a higher level of harvesting, and when the effect is destabilizing,  (\ref{ps2}), the sustainable level of harvesting is lower. In other words, if a system is more stable according to our susceptibility criteria, it would also be more stable according to the traditional definitions. Hence our conclusions about stabilizing (or dampening) and destabilizing effects of competition and intraguild predations are relevant for other definitions of stability. 

For our analysis, we considered topologically very simple systems of a linear food chains with a single additional link representing intraspecific competition or omnivory. In addition, we assumed linear functional responses to describe the predator-prey relation between adjacent trophic levels in the food chain. Nevertheless, our results can be generalized in a number of ways. 

\begin{figure}
\includegraphics[width=.7\textwidth]{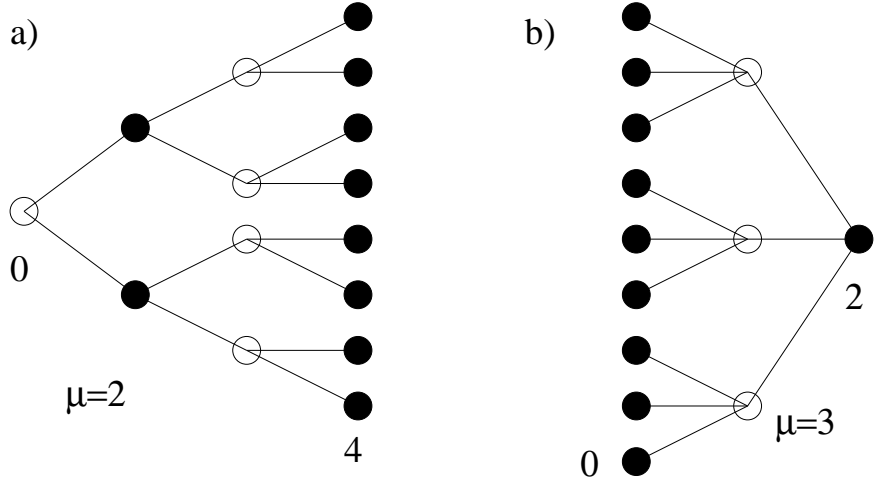}
\caption{\label{calfig}
Sketch of branching food webs (Cayley trees) in which $\m=2$ predators are feeding on the same  prey,  a), 
and in which each predator is feeding on $\m=3$ preys, b).}
\end{figure}

For example, numerical simulations reveal that our results remain true qualitatively with non-linear functional responses, as is shown in the Appendix. In Fig.~\ref{ash1b} we present an example of a destabilizing effect of an omnivory link, and in 
Fig.~\ref{ash2b} an example of stabilizing effect of omnivory is shown. It follows from these figures that in systems with Type-II functional response the effect of omnivory links
on stability is qualitatively similar to the structurally equivalent Lotka-Volterra counterparts.  
Moreover, our results can be generalized to more complicated forms of regular food webs, such as the Cayley trees illustrated in Fig.~\ref{calfig}. In ``upwards'' Cayley trees (Fig.~\ref{calfig} a), $m$ predators in the next higher trophic level are feeding on the same prey, and in ``downwards'' Cayley trees (Fig.~\ref{calfig} b), a given predator feeds on $m$ prey species on the next lower trophic level, so that for $m=1$ the linear food chains are recovered.

For example, for upwards Cayley trees (Fig.~\ref{calfig} a), the  population dynamics at the various trophic levels is given by 
\begin{equation}
\label{caley}
\frac{dx_k}{dt}=x_k\left(\l a x_{k-1} - \m a x_{k+1}- \d_k\right).
\end{equation}
In this case, the expression (\ref{full_solution}) for the steady state populations still holds,  but in all subsequent steps of the analysis presented in Section II, $\l$ should be replaced by $\l/\m$ and $d$ by $d/\m$. In the complimentary scenario of ``downwards'' branching (Fig.~\ref{calfig}b), $\l$ in  (\ref{full_solution}) and subsequent equations should be replaced by $\m \l$. Starting with these expressions, it is possible to derive results for the effects of self-interaction and omnivory on food web stability in Cayley trees, that are analogous to the results presented in Section III -- VI. In particular, omnivory and self-interaction can have both stabilizing and destabilizing effects in these more complicated food webs. Extrapolating from this, we conjecture that in more realistic food webs with more complex topologies and with a  multitude of self-interaction and omnivory links, it would be even harder to draw general conclusions about whether a particular link stabilizes or destabilizes the whole food web.

Our results may help to resolve the long-standing controversy on the effect of omnivory and self-interaction on the stability of food webs by essentially showing that general uniform conclusions regarding stabilization \cite{mccann97, holyoak98, fagan97,hillerislambers06} or destabilzation \cite{pimm77, holt97} cannot be drawn. Rather, whether a particular link in a food web leads to more or less stability very much depends on the details of how this link is embedded in the whole food web. For example, the sign of the effect depends on such features as the relative position of omnivory or self-interaction and the perturbation level from the top of the food web. In our simplified modeling situation, these positions are unambiguously defined. However, in reality, the top predator is often rare (e.g. big carnivores, which may often have low population densities), which will make the determination of positions relative to the top level difficult in real food webs. Nevertheless, it seems feasible that our results could be tested experimentally in simple and well-compartmentalized trophic systems.

\section {Appendix}
\subsection{Type-II functional response}
Here we show that the results derived above for the Lotka-Volterra type of functional response qualitatively hold for a more realistic, Type-II functional response. 
We consider a linear food chain with nearest neighbour predator-prey interactions  so that a rate of change of population density $x_k(t)$ of the $k$th trophic level is given by
\begin{equation}
\label{ai}
\frac{dx_k}{dt}=x_k\left[\frac{\l_ka_{k,k-1}x_{k-1}}{1+a_{k,k-1}h_{k} x_{k-1}} - 
\frac{a_{k+1,k} x_{k+1}}{1+a_{k+1,k}h_{k+1} x_{k}} - \d_k\right].
\end{equation}
Here $a_{i,j}$ are predation strength of species $i$ on species $j$, $\l_k$ is the conversion efficiency which connects the birth rate for the species $k$ to the amount of other species it consumes, $h_k$ is the handling time for species $k$, and
$d_k$ is the death rate coefficient of the species $k$.
Evidently, the top predator, occupying the $n$th level, does not have any species that prey on it, which could be expressed by setting $x_{n+1}\equiv 0$
In addition, the basal species, occupying the basal trophic level, is characterized by
the logistic growth term, which mimics the finite input of the energy into the system
(or finite carrying capacity $K$).
The linear  death term $d_0$ is absorbed into the linear part of the birth term $\b$, which gives for the rate of change of the basal species concentration, 
\begin{equation}
\label{a0}
\frac{dx_0}{dt}=x_0\left[\b\left(1-\frac{x_0}{K}\right) -  
\frac{a_{1,0}x_{1}}{1+a_{1,0}h_{1} x_{0}}\right].
\end{equation}

 As in the main text, we consider the effect of ``shortcuts'', or omnivory food links, reflecting predation of the species $u$ not only on its nearest neighbour $u-1$ below it in the food chain, but also on the species $w$ with the intensity $\a$ and conversion efficiency $\c$ (which replaces $\l$).
Introduction of such link modifies the rate equations (\ref{ai},\ref{a0}) for the species $w$,
\begin{eqnarray}
\label{sh1}
\nonumber
\left. \frac{dx_{w}}{dt}\right |_{omn}=\frac{dx_{w}}{dt}\\
 - 
x_{w}\left[\frac{\a a x_{u}}{1+h_u(a_{u,u-1} x_{u-1}+\a a x_{w})}\right],
\end{eqnarray}
the species $u-1$
\begin{eqnarray}
\label{sh2}
\nonumber
\left.  \frac{dx_{u-1}}{dt} \right |_{omn}=\frac{dx_{u-1}}{dt} +
a_{u,u-1}x_{u}x_{u-1}\left[\frac{1}{1+a_{u,u-1}h_{u} x_{u-1}}\right.\\
\left. - \frac{1}{1+h_u(\a  a x_{w} + a_{u,u-1} x_{u-1})} \right],
\end{eqnarray}
and for the species $u$
\begin{eqnarray}
\label{sh3}
\nonumber
\left.  \frac{dx_{u}}{dt} \right |_{omn}=\frac{dx_{u}}{dt} +
x_{u}\left[ - \frac{a_{u,u-1} \l_u x_{u-1}}{1+a_{u,u-1}h_{u} x_{u-1}} \right.\\
\left.  + \frac{\l_u a_{u,u-1}x_{u-1} + 
\c \a a x_{w} }{1+ h_u(a_{u,u-1} x_{u-1} + \a  a x_{w})} \right].
\end{eqnarray}

\begin{figure}
\includegraphics[width=.5\textwidth]{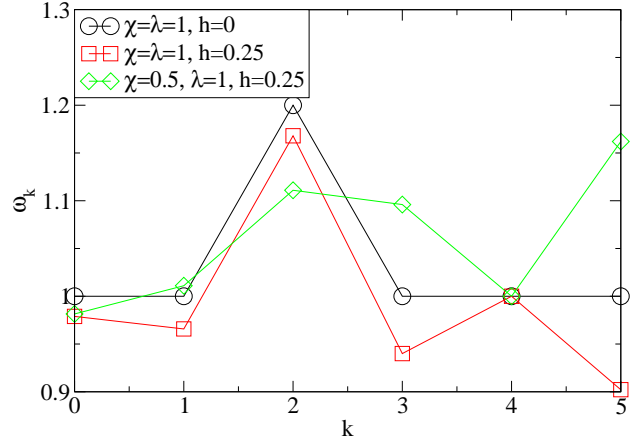}
\caption{\label{ash1b}
The ratio $\omega_k \equiv \frac {\left( x_k' - x_k\right)_{omn}}{x_k' - x_k}$ of the differences between the perturbed and unperturbed population with ($\a=0.2$) and without ($\a=0$) omnivory,
for Lotka-Volterra functional response  (circles) and Type-II functional response
with $h=0.25$ and $\c=\l=1$ (squares) and  $\c=0.5,\; \l=1$ (diamonds).}.  
\end{figure}

We use the following values of the constants, $a_{ij}=a=1$, $\l_i=0.75, \; 1$, $\c_i=0.5,\; 0.75, \; 1$, $K=1$, $\b=1$,
$\d_i=0.1$, and $h_i=0.25$ for all $i$ and $j$. As an example of destabililizing effect of intraguild predation, Fig.~\ref{ash1b}, we consider a linear food chain with an omnivory link of the intensity $\a=.2$ between levels $u=3$ and $w=1$. The  perturbation is caused by a shift by $\e a$, $\e=0.1$ of the death rate of the level $p=5$; all these 3 levels $\{u,w,p\}$  are in $\{n\}$ partition.

\begin{figure}
\includegraphics[width=.5\textwidth]{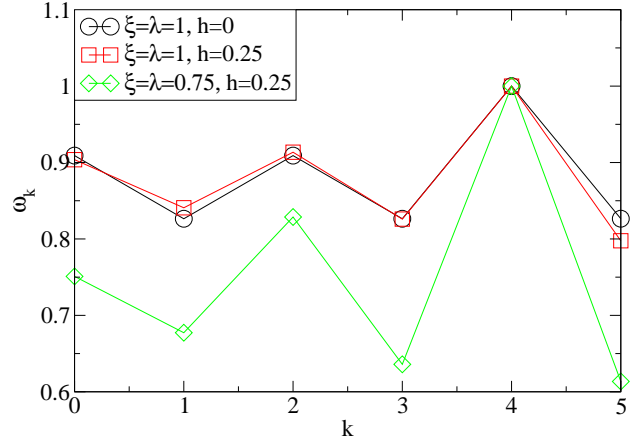}
\caption{\label{ash2b}
The ratio $\omega_k \equiv \frac {\left( x_k' - x_k\right)_{omn}}{x_k' - x_k}$ of the differences between the perturbed and unperturbed population with ($\a=0.1$) and without ($\a=0$) omnivory,
for Lotka-Volterra functional response  (circles) and Type-II functional response
with $h=0.25$ and $\c=\l=1$ (squares) and  $\c=l=0.25$ (diamonds).}.  
\end{figure}

As an example of stabilizing effect of intraguild predation, Fig.~\ref{ash2b}, we consider
an omnivory link between levels $u=3$ of the $\{n\}$ partition and $w=0$ of the $\{n-1\}$ partition.
The  perturbation is caused by a shift by $\e a$, $\e=0.1$ of the death rate of the level $p=5$; which is  in $\{n\}$ partition.

It follows from Figs.~\ref{ash1b}, \ref{ash2b} that the conclusions made in Section \ref{os} qualitatively hold for foodwebs with Type-II functional response as well: Depending on the position of the level with  a death rate shift and a shortcut link, the omnivory can either stabilize of destabilize
the food web. 
When  the omnivory link connects the levels $u=3$ and $w=1$  of the $\{n\}$ partition and the death rate shift occurs at the level $p=5$ also belonging to the $\{n\}$ partition, the perturbation is enhanced, especially at the level(s) $w<i<u$ of the  $\{n-1\}$ partition. Conversely, when the omnivory link connects the  levels $u=3$ of the $\{n\}$ partition and $w=0$ which belongs to $\{n-1\}$ partition and the shift of the death rate occurs at the level $p=5$ which belongs to the $\{n\}$ partition, the perturbation is dampened.

\begin{acknowledgments}
\end{acknowledgments}
\bibliography{verynew}
\bibliographystyle{plain}

\end{document}